# Ten Pillars for Data Meshes

Robert L. Grossman[1]*, Ceilyn Boyd[2], Nhan Do[3], Danne C. Elbers[3], Michael S. Fitzsimons[1], Maryellen L. Giger[1], Anthony Juehne[4], Brienna Larrick[1], Jerry S. H. Lee[5], Dawei Lin[6], Michael Lukowski[1], James D. Myers[2], L. Philip Schumm[1] and Aarti Venkat[1]

November 4, 2024

[1] University of Chicago
[2] Harvard University
[3] VA Boston Healthcare System
[4] HEAL Initiative/National Institutes of Health
[5] Ellison Institute of Technology
[6] NIAID/National Institutes of Health

*Corresponding author: Robert L. Grossman, University of Chicago, rgrossman1@uchicago.edu.

# Abstract

Over the past few years, a growing number of data platforms have emerged, including data commons, data repositories, and databases containing biomedical, environmental, social determinants of health and other data relevant to improving health outcomes.  With the growing number of data platforms, interoperating multiple data platforms to form data meshes, data fabrics and other types of data ecosystems reduces data silos, expands data use, and increases the potential for new discoveries. In this paper, we introduce ten principles, which we call pillars, for data meshes.  The goals of the principles are 1) to make it easier, faster, and more uniform to set up a data mesh from multiple data platforms; and, 2) to make it easier, faster, and more uniform, for a data platform to join one or more data meshes.  The hope is that the greater availability of data through data meshes will accelerate research and that the greater uniformity of meshes will lower the cost of developing meshes and connecting a data platform to them.

# Introduction

Over the past few years, a growing number of data platforms have emerged, including data commons, data repositories, and databases containing biomedical, environmental, social determinants of health and other data relevant to improving health outcomes.

With the growing number of data platforms, interoperating multiple data platforms to form data meshes, data fabrics and other types of data ecosystems reduces data silos, expands data use, and increases the potential for new discoveries.

In this paper, we introduce ten principles, which we call pillars, for data meshes.  The goals of the principles are 1) to make it easier, faster, and more uniform to set up a data mesh from multiple data platforms; and, 2) to make it easier, faster, and more uniform, for a data platform to join one or more data meshes.  The hope is that the greater availability of data through data meshes will accelerate research and that the greater uniformity of meshes will lower the cost of developing meshes and connecting a data platform to them.

The principles are divided as follows: four pillars apply to a data platform that wants to join a data mesh; five pillars apply for setting up and operating a data mesh; and, one pillar applies to data platforms that provide analysis environments for meshes.

When there is some level of consensus, we reference standards and identify community best practices.

We use the following terminology in this paper, following [1].

We use the term *data platform* as a general term for a software platform that includes data repositories, data commons, databases, knowledge repositories, knowledgebases and other software platforms for working with data.

A *data mesh* contains multiple data platforms and other applications for analyzing and sharing data called *nodes* that can interoperate using a common set of software services, which are called *mesh services*.  Typically mesh services include software services for authentication, authorization, accessing data, linking data, and related functions.

A typical mesh contains two or more data platforms providing data, such as data repositories or data commons, a *data hub* enabling a user to search for particular data or datasets, and one or more *authorized analysis environments* (aka trusted analysis environments).  Usually, in a data mesh, the data remains in the data platforms that host the data, the hub provides a means to search and discover data, and the data is analyzed in an authorized analysis environment.  The authorized analysis environment may be attached to the data platform that hosts the data, attached to another data platform in the mesh, or may be managed by the user's organization.

# Mesh Data Objects and Their Metadata

As just mentioned, in a data mesh, we assume that data objects are managed by the data platforms that host them.  Typically, the metadata about these data objects are made available to the data mesh so that data hubs, or other data platforms in the data mesh, can discover data objects of interest, and analyze them in approved authorized analysis environments (cf. [2]).

As part of the data mesh governance, the data mesh typically agrees on:

1. **What types of *data objects* are being indexed and made available for access and analysis by the mesh.** Common examples include: datasets, research studies that may contain one or more datasets, clinical trials, research participants or patients and their associated clinical data, research participants or patients and their associated omics data, research participants or patients and their associated imaging data, research participants or patients and their associated multi-modal data, BAM files and associated clinical and other data, DICOM objects and associated clinical data and other data.  See Pillar 6 below.

2. **The minimum metadata required for each data object managed by the mesh.**  In this article, we use the term *Data Mesh Metadata (DMM)* for the metadata managed by the data mesh.  This helps distinguish this metadata from the metadata associated with the data object that is managed by the data platform that hosts the data object. In general, we assume that the DMM is public, but there is nothing to preclude data meshes supporting both private and public DMM, or entirely private DMM.  See Pillar 7 below.

3. **Security and compliance requirements.** A data mesh may also specify minimum security and compliance requirements for a data platform to join the mesh and for a data platform to be a trusted analysis environment for the mesh.

4. **Services to link research subjects or patients across two or more data platforms in the data mesh.**  Examples include privacy preserving record linkage services (PPRL) with honest brokers and PPRL with globally unique IDs (GUIDs) that are created as data is contributed to data platforms.

Notice that a particular data mesh may index multiple types of data objects.  For example, a data mesh may index both research studies in general and particular types of research studies, such as clinical trials, and decide that the minimum metadata required for these two types of data objects are different.

# General Pillars so that a Data Platform Can Participate in a Data Mesh.

Recall that we are using the term data platform, which is a node in the data mesh, to refer to a data commons, data repository, or software platform that contains data and makes it available to a community.

**Pillar 1.** Data in a data platform should be identified with Persistent Identifiers (PIDs).

The most familiar example of a PID are DOIs assigned to datasets [3]. Another important type of persistent identifiers are systems that assign GUIDs or UUIDs to data objects [4], such as data GUIDs [5] or ARK identifiers [6]. As an example, an image series in a data commons may be assigned a DOI, while each image slice in the series may be assigned a GUID managed by the data commons or associated data mesh.

Note that Pillar 8 below distinguishes between PIDs issued by the data repository and PIDs issued by the data mesh.

**Pillar 2.** Data in a data platform should have FAIR APIs so that the metadata associated with a PID can be accessed.

Note that using the metadata API of a data platform in the mesh, a data hub in a data mesh can collect the metadata required to support search and discovery across the data mesh.

It's a best practice in a data mesh that enough of the metadata should be public so that search across a mesh can be supported.

**Pillar 3.** Data in a data platform should have FAIR APIs so that the data associated with a PID can be accessed.

Best practices for accessing biomedical data via FAIR APIs include the use of the GA4GH DRS standard [7] and signposting [8].

It is important to note that the data repository may restrict the *transfer of data* in a data mesh to research environments that it has individually evaluated and made the determination that the analysis environment is authorized to support the analysis of data by researchers that are authorized to access the data. See Pillar 5 below.

**Pillar 4.** There should be an API for authentication and authorization.

This pillar is critical so that data hubs in a data mesh can identify which data is available to a user and so that data can be transferred when requested by the user to an appropriate authorized analysis environment when the user is authorized to access the data.

The authentication API should preferably use OpenID Connect. One standard for authorization that is growing in acceptance is the GA4GH standard for visa and passports [7]

# Pillars for Connecting to Authorized Analysis Environments

**Pillar 5.** A data platform in a data mesh should provide an API so that authorized analysis environments can access data on behalf of authenticated users who are authorized to access data in the platform [2].

Note that Pillar 2 provides metadata so that datasets in a mesh can be discovered, while Pillar 3 makes a dataset in a mesh accessible by an authorized analysis environment attached to a hub. Also note that in some cases, analysis can be done using open access metadata and data in any analysis environment, not just an authorized analysis environment.

A best practice is for each dataset provided by the data repository to the authorized analysis environment to specify whether the dataset can be downloaded out of the authorized analysis environment, can be redistributed, etc following [2].

It's important to note that even with this API, the data repository has the sole authority to determine whether a given analysis environment is authorized or not.

# Pillars for Forming a Data Mesh

There are a wide range of data meshes, from formal data meshes developed and operated by a data mesh sponsor to informal data meshes that arise by using common standards, APIs, and minimum metadata models. Particular data meshes may decide to specify or leave unspecified decisions about standards, APIs, data mesh consortium governance, data governance, platform governance, etc. This results in a spectrum in each of these and other dimensions ranging from less governed, more informal meshes to more governed, more formal meshes. See Figure 1.

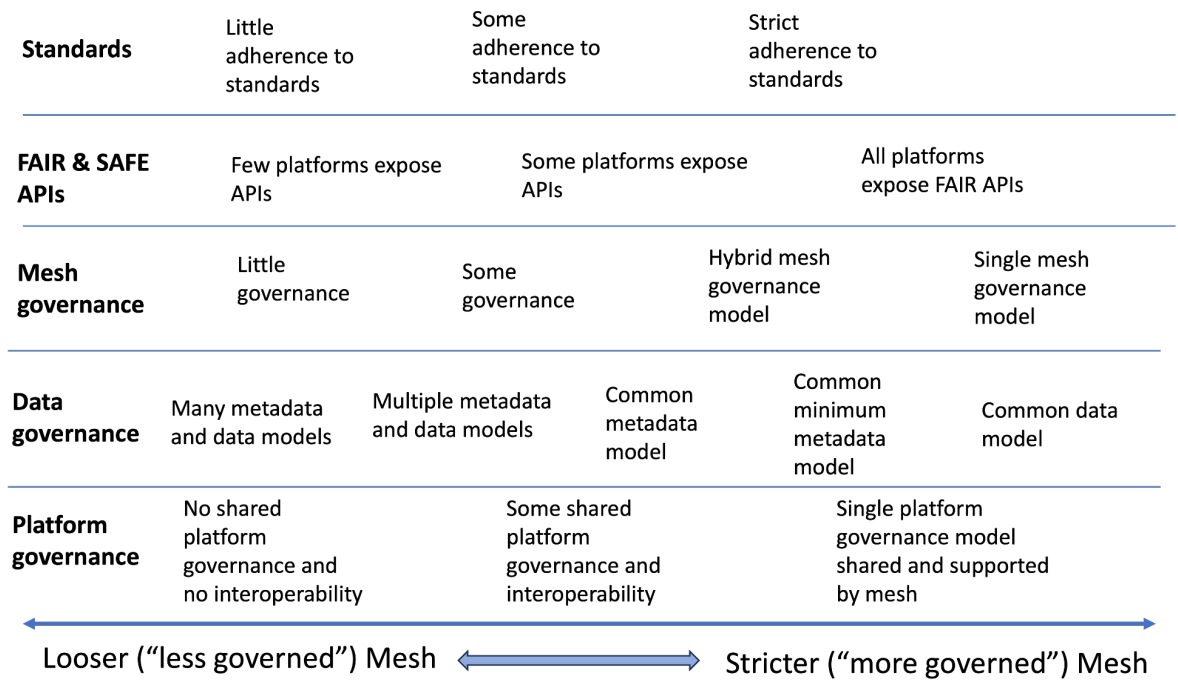

| Standards | Little adherence to standards | Some adherence to standards | | Strict adherence to standards | |
|---|---|---|---|---|---|
| FAIR & SAFE APIs | Few platforms expose APIs | Some platforms expose APIs | | All platforms expose FAIR APIs | |
| Mesh governance | Little governance | Some governance | Hybrid mesh governance model | | Single mesh governance model |
| Data governance | Many metadata and data models | Multiple metadata and data models | Common metadata model | Common minimum metadata model | Common data model |
| Platform governance | No shared platform governance and no interoperability | Some shared platform governance and interoperability | | Single platform governance model shared and supported by mesh | |

Looser ("less governed") Mesh ⟷ Stricter ("more governed") Mesh

Figure 1. There is a wide range of decisions that data meshes make along multiple dimensions, including standards, APIs, mesh governance, data governance and platform governance, leading to less governed, more informal data meshes to more governed, more formal meshes.

A particular data mesh should specify a shared governance model that covers data governance [9], platform governance [2], and sustainability that is suitable for their mesh. Shared governance in this context spans the repositories or commons participating in the mesh as well as the hubs supporting search and discovery and the connected workspaces, analysis environments, and computing infrastructure for exploring and analyzing the data.

It is important to keep in mind that the node in the mesh that hosts the data has primary responsibility for data governance. As an example, access to data for a user (authorization) is controlled by the node in the mesh that hosts the data. It's a matter of platform governance [2] whether the node that hosts the data can transfer the data to other nodes in the mesh with the required security and compliance, including analysis environments. Challenging issues such as negotiating Data Use Agreements (DUA) [10] do not fundamentally change when a node is part of a mesh. Of course, an overly restrictive DUA may preclude data being accessed by an authorized user from other platforms in the mesh that have the required security and compliance. This is one of the important roles of platform governance.

Note that unlike a data repository that needs a long sustainability plan for the data it stewards [11], a data mesh may be structured for a finite period of time, say to support a 3 year, 5 year or 10 year project, as well as for longer terms.

Pillars 6-10 below are relevant to data meshes with governance models. As mentioned, data meshes may be informal and arise when different data platforms use compatible APIs and standards or may be more formal with some level of data mesh governance. Note that data mesh governance is usually a hybrid governance model, with each data platform or resource in the mesh having its own organizational governance structure and with shared governance in some areas, such as mesh services, minimum metadata, etc.

**Pillar 6.** A data mesh should agree on a shared governance model covering data governance and platform governance.

Note that there is a range of shared governance in data meshes. For some data meshes, the shared governance carefully and completely covers data governance and platform governance (see [12] for an example), while for other data meshes the shared governance is minimal and the data hub simply needs to check carefully that its data governance is consistent with the required data governance for each dataset in each data platform participating in the data mesh. Data mesh governance may also cover sustainability. Data mesh governance sometimes includes key components of a consortium governance for the data mesh beyond data and platform governance or reference separate consortium governance agreements.

**Pillar 7.** A data mesh should agree on the types of data objects supported, and the *minimum* metadata required for each type of data object.

We call the metadata associated with data objects in the mesh the Data Mesh Metadata (DMM) or more precisely the DMM for data objects.

For example, a data mesh may support research studies, including clinical trials as the basic data objects supported in the data mesh. In this case, the data mesh would specify as part of the data mesh data governance the minimum metadata required for research studies and clinical trials that contribute data to data platforms in the data mesh.

In some cases, some of the minimum metadata may not be available until after a dataset is added to a data platform or may not be available from the data platform's API. In these cases, the data hub may acquire this metadata from the data contributor and/or other sources. This additional metadata will be imported to the Data Mesh Metadata Service (DMMS).

**Pillar 8.** The DMMS should assign a PID to each object supported. For example, it should assign a PID to each dataset, study or other entity in the mesh. The metadata associated with this PID should specify the data platform that hosts the data and the primary PID assigned by the data repository or commons hosting with the entity. It may also reference the DOIs of the primary publication(s) associated with the entity.

Entity as used here often refers to a datasets or studies, but may refer to other entities also, such as patients or research subjects with genomic data (e.g. [13]), patents or research subjects with image data (e.g. [14]), etc.

Note that although some data repository architectures and policies may limit the ability to assign a PID to each data file or data object and may just assign PIDs to datasets, the data mesh may choose to assign PIDs at a finer level of granularity.

**Pillar 9.** Return of usage statistics. Each data hub in the data mesh should return usage statistics and usage information of who used the data hub to access data to the data repository that provides the data.

Return of usage statistics is optional for a data mesh, since some nodes will not join a mesh if this is not provided, while others may not join a mesh if it is required.

Note that since different commons/repositories have different requirements for data access, such as approval by a data access committee, registration, or completely open access with no registration required, the type of information returned will vary. For example, for completely open access data without any registration, only counts can be returned; while for registered or controlled access data, names of those individuals accessing the data can be returned.

The data hub should also inform the user that such data is being collected through a pop up window or other mechanism.

**Pillar 10.** The data mesh operator should provide a public API interface to the DMMS for the data mesh metadata (DMM).

As usual, the API should be designed in such a way so that the DMM is FAIR. In particular, a permissive license should be associated with the DMM so that it can be freely used to create other data ecosystems.

# Some Special Types of Data Meshes

**Data meshes supporting search, access, and analysis.** A very important type of data mesh is a data mesh containing a data hub that supports search and discovery over two or more data platforms in the data mesh. A best practice is for the data hub to also support one or more authorized analysis environments that are authorized to analyze data [2] from the data platforms in the data mesh. This type of data mesh also agrees on the types of data objects it will support and on the minimum metadata for each. It is important to note that the original data always remains in the associated data repository and that the data mesh returns usage statistics to the associated data repository.

**Data meshes with registration services**.  The data mesh operator for some data meshes can provide a registration service so that data in a data repository (or that will be deposited in a data repository in the future) can be registered with data mesh.  Both a data registration portal and a data registration API should be supported so that a user with a dataset can register the dataset.

**Data meshes supporting federated analysis.**   Another type of data mesh contains two or more data platforms and services so that computations can be moved to the data, instead of moving the data to a centralized computing platform.  A data mesh containing two or more trusted research environments or data enclaves can support the federated analysis of data in this way.  For example, a data mesh of this type might require as part of its governance the following requirement:

> A data platform in a mesh supporting federated queries or computations should provide an API so that approved workflows can be submitted to the data platform in the data mesh, executed over the indicated data in the data platforms, and the results returned to the user, possibly after review.

# Acknowledgments

Research reported in this publication was supported in part by the National Institute of Biomedical Imaging and Bioengineering (NIBIB) of the National Institutes of Health under contract 75N92020D00021/5N92023F00002 (MLG and RLG); by the National Institutes of Health HEAL Initiative under contract 3OT2OD030208-01S3 (LPS and RLG); and by the NSF under Grant #2116935 (JDM).  The views and conclusions contained in this document are those of the authors and should not be interpreted as representing the official policies, either expressed or implied, of the NIH or NSF.

# Conflicts of Interest

The authors have no conflicts to declare.